\documentclass[final,3p,times,twocolumn]{elsarticle}

\usepackage{graphics}

\usepackage{amssymb}

\journal{Solid State Communications}

\begin{document}

\begin{frontmatter}



\title{Evidence for magnetic phase separation in colossal magnetoresistance compound EuB$_{5.99}$C$_{0.01}$}

\address[label1]{Institute  of  Experimental  Physics,
 Slovak   Academy  of Sciences, Watsonova 47,
 040~01~Ko\v {s}ice, Slovakia}
\address[label2]{Polish Academy of Sciences, Institute of Low Temperature and Structure Research, 50-422 Wroclaw, Poland}
\address[label3]{Helmholtz Zentrum Berlin, D-14109 Berlin, Germany}
\address[label4]{Institute for Problems of Material Science, NASU, 252680~Kiev, Ukraine}

\author[label1]{I. Batko} 

\author[label1]{M. Batkova \corref{cor1}}
\cortext[cor1]{E-mail: batkova@saske.sk; Fax: +421 55 6336292; Telephone: +421 55 7922239}

\author[label2]{V. H. Tran} 

\author[label3]{U. Keiderling} 

\author[label4]{V. B. Filipov}

\address{}

\begin{abstract}
EuB$_{5.99}$C$_{0.01}$ is a low-carrier density ferromagnet that is believed to be intrinsically inhomogeneous due to fluctuations of carbon content. In accordance with our previous studies, electric trasport of EuB$_{5.99}$C$_{0.01}$ close above temperature of the bulk ferromagnetic (FM) ordering is governed by magnetic polarons. 
Carbon-rich regions are incompatible with FM phase and therefore they act as spacers preventing magnetic polarons to link, to form FM clusters, and eventually to percolate and establish a (homogoneous) bulk FM state in this compound, what consequently causes additional (magneto)resistance increase. Below the temperature of the bulk FM ordering, carbon-rich regions give rise to helimagnetic domains, which are responsible for an additional scattering term in the electrical resistivity. Unfortunately, there has not been provided any direct evidence for magnetic phase separation in EuB$_{5.99}$C$_{0.01}$ yet. 
Here reported results of electrical, heat capacity, Hall resistivity and small-angle neutron scattering studies bring evidence for formation of mixed magnetic structure, and provide consistent support for the previously proposed scenario of the magnetoresistance enhancement in EuB$_{5.99}$C$_{0.01}$.  
\end{abstract}


\end{frontmatter}

\section{Introduction}
\label{Intr}

EuB$_6$ is a rare example of low carrier density hexaboride that orders ferromagnetically 
			at low temperatures via two consecutive phase transitions \cite{Degiorgi97}. 
Physical properties of this system are thought to be governed by magnetic polarons (MPs)
		 \cite{Snow01,Sullow00a,Calderon04,Yu06,Das012}. 
As indicated by Raman-scattering measurements \cite{Snow01}, MPs have set in at about 30 K
		 when cooling EuB$_6$ from a higher temperature.
According to S\"ullow et al. \cite{Sullow00a}, the magnetic phase transition at 15.5 K 
			represents emergence of spontaneous magnetization accompanied with metallization. 
At this temperature MPs begin to overlap and form a conducting, ferromagnetically ordered 
			phase that acts as a percolating, 
			low-resistance path across the otherwise poorly conducting sample \cite{Sullow00a, Das012}. 
With decreasing temperature the volume fraction of this conducting ferromagnetic (FM) phase expands
		until the sample becomes a homogeneous conducting bulk ferromagnet at 12.6 K \cite{Sullow00a}. 

High-pressure measurements indicate that the FM order is driven by an RKKY interaction 
		between the localized Eu moments and the very dilute pocket of conduction electrons
			 arising from a semi-metallic band overlap \cite{Cooley97}.
Due to a very low number 
			 of intrinsic charge carriers,~10$^{20}$ cm$^{-3}$ \cite{Aronson99},
			  the system is very sensitive even to a slight change in concentration of conduction electrons,
			   e. g. due to change in chemical composition or in impurity concentration \cite{Kasaya78,Molnar81}. 
A substitution of boron by carbon in EuB$_6$ increases number of conduction electrons 
		in the conduction band of EuB$_6$ \cite{Etourneau85}, 
			thus EuB$_{6-x}$C$_{x}$ carbide borides behave as degenerate semiconductors, 
			in which both, carrier concentration, and antiferromagnetic  
		interaction increase with increasing carbon content \cite{Etourneau85}. 
As it was shown by neutron diffraction studies, 
		EuB$_6$ behaves like a simple ferromagnet, whereas EuB$_{5.80}$C$_{0.20}$ has an incommensurate
		spiral structure \cite{Etourneau85,Tarascon81}, 
		and the magnetic structure of intermediate EuB$_{5.95}$C$_{0.05}$  can be described 
		as a mixture of FM and helimagnetic (HM) domains \cite{Etourneau85,Tarascon81}. 
Appearance of the HM domains is associated with local increase of carbon concentration in the material. 
Size of these incoherent regions is estimated to be about 5~nm \cite{Tarascon81}. 
The presence of the HM domains formed in carbon-rich regions due to distinct 
			impact of the RKKY interaction 
		because of distinct carrier density \cite{Tarascon81} is believed to be responsible 
		for an additional scattering term in the electrical resistivity \cite{Batko95}.  

Electric, magnetic, and heat capacity studies of EuB$_{5.99}$C$_{0.01}$ support the hypothesis 
			that the dominant scattering process in this material at temperatures below the bulk magnetic 
			transition, $T_C = 4.3$~K \cite{Batkova08}, has its origin in the mixed magnetic structure 
			\cite{Batko95,Batkova08, Batkova10}. 
The anomalous transport properties of the EuB$_{5.99}$C$_{0.01}$ can be satisfactorily explained 
		assuming the presence of MPs \cite{Batkova08}. 
It has been proposed that carbon-rich regions act as spacers, 
		which prohibit formation of a conducting, ferromagnetically ordered path across the sample.
As a consequence, the system persists in a poorly conducting state down to lower temperatures. 
Due to the extended temperature interval, in which the resistivity increases upon cooling,
		 an additional resistivity increase is observed, resulting in the higher value of the resistivity maximum.
Such a scenario, assuming the presence of MPs, allows also to explain why the resistivity maximum 
		of the EuB$_{5.99}$C$_{0.01}$
    ($\approx390~\mu\Omega$.cm at $\approx5$~K) is larger than that of the stoichiometric EuB$_6$ 
    ($\approx350~\mu\Omega$.cm at~15~K),
     although the EuB$_{5.99}$C$_{0.01}$ is about four times better conductor at room temperature, 
     having $\rho(300 K)\approx180~\mu\Omega$.cm, 
     than the EuB$_6$ with  $\rho(300 K)\approx730~\mu\Omega$.cm. 
An important consequence of this scenario \cite{Batkova08} 
     is that it might show a route for an optimization of magnetoresistive properties also in other
      spatially inhomogeneous systems with magnetic polarons or with FM phase in general \cite{Batkova08}. 
Unfortunately, there has not been provided any direct evidence for magnetic phase separation 
		in EuB$_{5.99}$C$_{0.01}$ yet.
The purpose of this paper is to bring experimental support for the presence of magnetic phase 
		separation phenomena 
		in EuB$_{5.99}$C$_{0.01}$ based on performed Hall-effect, heat-capacity, 
		and small angle neutron scattering (SANS) studies.

\section{Material and Methods}
	
All samples studied in this work were cut from the same single crystal 
		grown by means of the zone floating used in our previous studies \cite{Batko95,Batkova08,Batkova10}.
The resistance, Hall resistance, heat capacity, and magnetisation were measured 
			in Physical Property Measurement System and Magnetic Property Measurement System (Quantum Design, USA). 
			SANS measurements were carried out at the V4 instrument at Helmholtz-Zentrum Berlin.

\section{Results and Discussion}	

\begin{figure}[!h]
\resizebox{1.0\columnwidth}{!}{%
  				\includegraphics{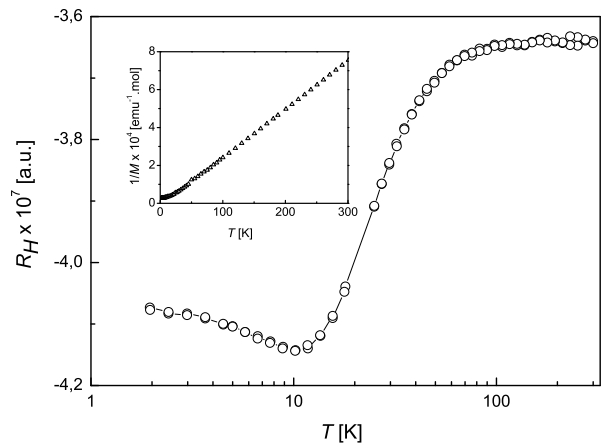}
  				}
\caption{Temperature dependence of the Hall resistivity ($R_H$) for EuB$_{5.99}$C$_{0.01}$ measured at 5~T 
				in the temperature range 2~-~300~K. Inset shows the corresponding inverse magnetic susceptibility 
				for the same value of magnetic field.}
\label{G-RH}
\end{figure}

Hall resistivity ($R_H$) and magnetization ($M$) measurements were performed 
		in the temperature range of $2 - 300$~K and in magnetic fields up to 5~T. 
The temperature dependences of $R_H$, and $1/M$ measured at the magnetic field of $B = 5$~T are shown in Fig.\ref{G-RH}. 

\begin{figure}[!t]
\resizebox{1.0\columnwidth}{!}{%
  				\includegraphics{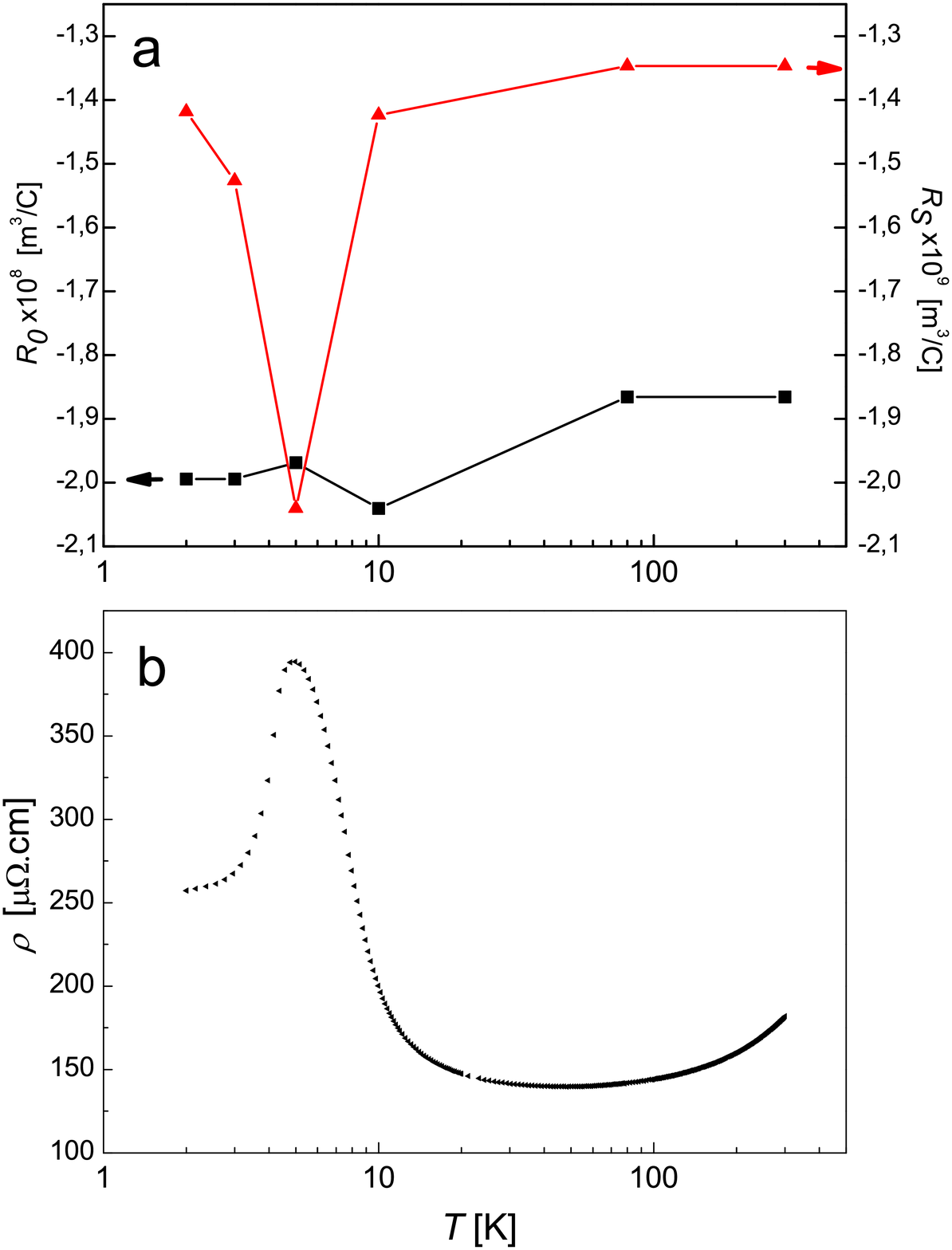}
  				}
\caption{Temperature dependences of a) $R_S$ and $R_0$,
and b) the electrical resistivity of EuB$_{5.99}$C$_{0.01}$.}
\label{G-R0}
\end{figure}

Hall resistivity in magnetic conductors can be expressed as $R_H = R_0 + 4\pi R_SM/B$, where $R_0$ and $R_S$
	    are the ordinary and anomalous Hall coefficients, respectively \cite{Hurd72}.
Therefore, $R_0$ and $R_S$ can be adequately determined when combining both, the Hall 
			resistivity and magnetization data.
Our analysis shows that neither  $R_S$ nor  $R_0$ changes with temperature in the temperature 
			range of $80 - 300$~K (see Fig.~\ref{G-R0}a). 
However, $R_S$ was found to change significantly at lower temperatures, especially around 
			the temperature of the FM ordering, $T_C = 4.3$~K [13], where it exhibits an extremum 
			as typically observed in magnetic materials.
Comparison of $R_S(T)$  to the $\rho(T)$ dependence (see Fig.~\ref{G-R0})
			reveals that processes governing 
				electric transport in region of the resistivity maximum are adequately sensed
				 in the anomalous term of the Hall resistivity. 
On the other hand, there is only weak change of $R_0$ in the whole temperature range. 
The value of $R_0$ at room temperature corresponds to the electron concentration of $3.3 x 10^{20}$~cm$^{-3}$, 
		which is approximately 3-times greater than $1.2 x 10^{20}$~cm$^{-3}$ reported for the stoichiometric EuB$_6$ \cite{Aronson99}. 
Such a temperature dependence of $R_0$ coincides with our expectations.
As showed in our previous studies of optical reflectivity of EuB$_{5.99}$C$_{0.01}$ \cite{Batko95}
			 the plasma edge frequency value, $\omega_p = (ne^2/\epsilon_0m_e)^{1/2}$ (here $n$ is the electron concentration, 
			 $e$ is the electron charge, $\epsilon_0$ is the vacuum permitivity, and $m_e$ is the electron mass) 
			 has been found to be almost constant between 4.5 and 30~K \cite{Batko95}. 
The finding infers that the electron concentration $n$ does not change substantially in the
		 whole temperature range studied \cite{Batko95}.
Thus, the Hall resistivity data obtained in this work bring an independent proof that strong 
		temperature dependence of the electrical resistivity of  EuB$_{5.99}$C$_{0.01}$ in the vicinity and close
		 above the temperature of the bulk FM ordering (see Fig.\ref{G-R0}b) is not a consequence of changes 
		 in electron concentration, but it is a result of scattering processes, presumably 
		 due to magnetic phase separation. 

	 		 \begin{figure}[!b]
\resizebox{1.0\columnwidth}{!}{%
  				\includegraphics{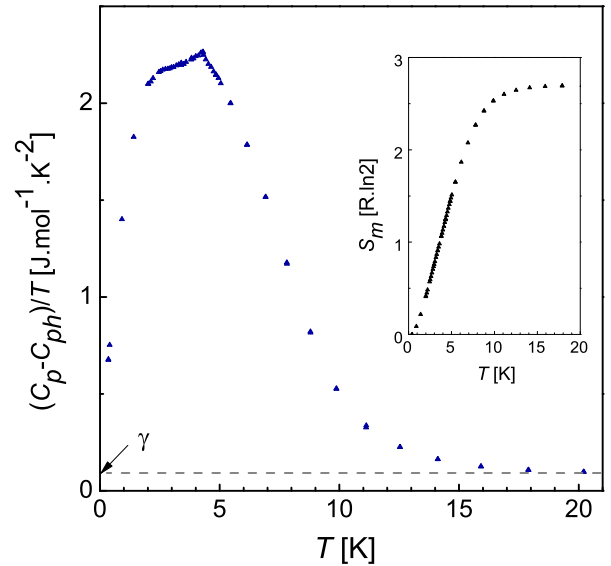}
  				}
\caption{Temperature dependence of (C$_p$ - C$_{ph}$)/T of EuB$_{5.99}$C$_{0.01}$ in the temperature region from 0.35 to  20 K. Inset shows calculated temperature dependence of the magnetic entropy $S_m$.}
\label{CpvsT}
\end{figure}	 

Heat capacity, $C_p$, of EuB$_{5.99}$C$_{0.01}$ and the isostructural non-magnetic compound LaB$_6$ was measured
		 in the temperature range $0.35 - 20$~K.
The phonon contribution, $C_{ph}$, for EuB$_{5.99}$C$_{0.01}$ was estimated 
			from the heat capacity of LaB$_6$.
As can be seen in Fig.~\ref{CpvsT}, the $(C_p - C_{ph})/T$ dependence of the studied system 
			shows a broadened continuous decrease in the temperature region between 5 and 15~K, 
			i.e. above the temperature of the bulk ferromagnetic ordering. 
This decrease can be ascribed
			 to a short range magnetic ordering. 
A clear evidence of the FM order appears as local maximum in the 
		$(C_p - C_{ph})/T$ curve at 4.3~K. 
It is remarkable that $(C_p - C_{ph})/T$ goes through a knee around 2.5 K,
		 implyimg the crystal electric field splitting of energy levels of the Eu$^{2+}$ ions.
$(C_p - C_{ph})/T$ drops rapidly with further decreasing temperature, but finally attains 
			a rather huge $(C_p - C_{ph})/T$ value of $\cong$0.5~Jmol$^{-1}$K$^{-2}$ at 0.35~K.  
Due to the nearby FM order transition, this value cannot be taken as an electronic specific heat coefficient. 
However, as temperature approaches 20~K, $(C_p - C_{ph})/T$ dependence becomes constant, 
			which thus represents the electronic specific heat only. 
Thus, for purposes of our analysis we have estimated the coefficient $\gamma$
		  to be the value $(C_p - C_{ph})/T$ at 20~K
		 (see Fig.~\ref{CpvsT}). 
Such estimation yields a $\gamma$ value of $\cong$0.1~Jmol$^{-1}$K$^{-2}$. 
In combination with the low charge-carrier concentration 
		$n \cong3.3 x 10^{20}$~cm$^{-3}$, the enhanced electronic specific heat could be associated either with 
		(i) an anomalously large effective mass of the electrons $\propto$190~$m_e$ (like in heavy fermions) 
		or (ii) a strong disorder of the system. 
Because of the nature of the system and according to SANS measurements discussed below we prefer the latter case.
                              
Magnetic contribution to the heat capacity, $C_m$, is assumed to be $C_m = C_p - C_{ph} - \gamma T$. 
The temperature dependence of the magnetic entropy, 		
		$S_m(T) = \int^{20K}_0{\frac{C_m(T)}{T}}dT$,
was calculated and shown in the inset of Fig.~\ref{CpvsT}. 
As can be seen the magnetic entropy of the EuB$_{5.99}$C$_{0.01}$ at 20~K reaches about 2.7~Rln2, 
			which is comparable to the theoretical entropy of the magnetic ordering of Eu~$^{2+}$ ions. 
A similar reduction in $S_m$ was observed also for the stoichiometric EuB$_6$ \cite{Sullow00a} .  

Small-angle neutron scattering (SANS) experiments were performed at temperatures between 2 and 30~K 
			with the aim to provide direct evidence for presence of magnetic phase separation in the
			 temperature range of 2~-~7~K.
Considering the physical state at 30~K as a reference paramagnetic state without 
			magnetic phase separation \cite{Snow01}     
		 we focused on evolution of magnetic state in the mentioned temperature interval.
The SANS measurements were performed for scattering vectors Q from 
			the interval of $0.3 - 3.4$~nm$^{-1}$. 
Supposing the simplest case of scattering due to spherical objects, such interval enables to detect objects 
			having size/diameter between ~1.8 and ~20~nm. 
Here should be noted that our earlier size estimation of the "spacers" in EuB$_{5.99}$C$_{0.01}$ 
			has provided the value ~2.9~nm \cite{Batkova10}. 

\begin{figure}[!t]
\resizebox{1.0\columnwidth}{!}{%
  				\includegraphics{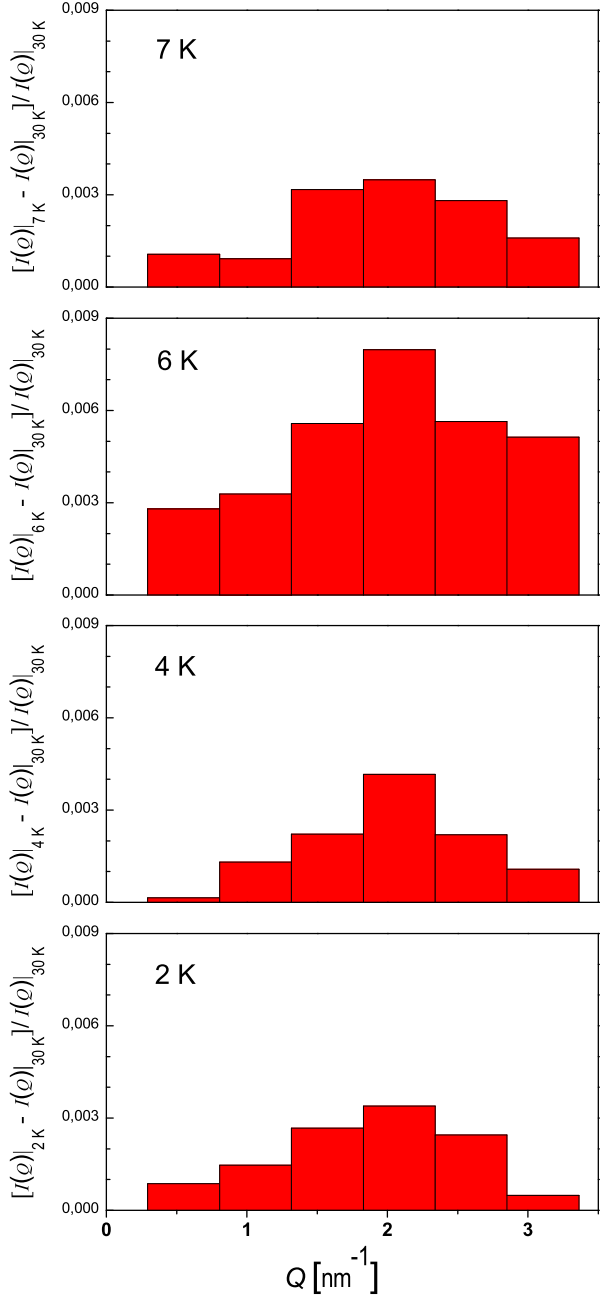}
  				}
\caption{Averaged $(I(Q)|_T-I(Q)|_{30~\scriptsize\textnormal{K}})/I(Q)|_{30~\scriptsize\textnormal{K}}$
			 dependences calculated from SANS data collected 
			at temperatures 2~K, 4~K, 6~K, 7~K, and 30 K (see text).}
\label{SANS}
\end{figure}

The usual way for the interpretation of the scattered neutron intensity I(Q) would be: 
			(1) correction of the data for experimental parameters like sample transmission,
			 background scattering and local detector efficiencies, and then 
			 (2) calculation of a particle size distribution, the scattering of which would
			  produce just this corrected scattered neutron intensity. 
The structural  changes in the sample as a function of the varied experimental parameter 
			(in this case the temperature) would then become directly obvious when the different
			 particle size distributions are compared. 
However, the studied sample was prepared  from natural boron, which contains approximately 
				20$\%$ of highly neutron-absorptive isotope of $^{10}$B. 
This lead to an extremely low neutron transmission of the sample of less than 1$\%$. 
As a consequence, the intensity scattered from the sample itself was so low and noisy 
		that it was not possible to perform the usual data correction procedure to separate 
		the actual sample scattering from the background contributions.
We even noticed that the scattered intensity was sensitive to very small 
		displacements of the unit of sample holder plus sample in the beam, 
		indicating that the scattering contribution from the sample holder 
		was even higher than the contribution from the sample. 
Therefore, we chose the following alternative way to conduct and interpret the experiment: 

Measurements at all temperatures were done without moving the sample inbetween. 
Since no separate background scattering could be measured, all
			scattering intensities remain a mixture of sample and background scattering. 
This means that no calculation of explicit size disributions is possible, 
			and only qualitative interpretation can be done directly on the scattering curves. 
These interpretations are based on the fact that particles with a diameter $d$ produce 
			a scattering contribution in the range order of $Q \cong 2\pi/d$. 
From the known properties of the used cryomagnet we could assume that the background contribution 
		would not change within the investigated temperature range. 
To eliminate the very large contribution of the background, we considered the relative 
		change of the intensities $(I(Q)|_T-I(Q)|_{30~\scriptsize\textnormal{K}})/I(Q)|_{30~\scriptsize\textnormal{K}}$. 
Since we know that at $T = 30$~K none of the expected scattering objects should be present, 
		this relative comparision of the scattering at a temperature $T$ and at the reference
		 temperature $T = 30$~K  will extract the very small intensity contribution added by those objects. 
To reduce the noise further, we divided the investigated $Q$ range into only 6 intervals, 
		so that each interval was an average over approximately 9 detector cells. 
This of course significantly reduces the $Q$ 
			resolution and therefore the sensitivity of the experiment to small changes 
			in the number and size of the scattering objects. 
However, this approach is sufficient to detect the qualitative results, 
		that are discussed in the following.
This way averaged $(I(Q)|_T-I(Q)|_{30~\scriptsize\textnormal{K}})/I(Q)|_{30~\scriptsize\textnormal{K}}$
			dependences for temperatures 2~K, 4~K, 6~K, and 7~K 
			are shown in Fig.~\ref{SANS}. 
	
As can be seen in Fig.~\ref{SANS}, temperature changes in the vicinity of the temperature of the resistivity maximum ($\approx$6~K)
		 reveal clear evolution in the $(I(Q)|_T-I(Q)|_{30~\scriptsize\textnormal{K}})/I(Q)|_{30~\scriptsize\textnormal{K}}$
		  dependences, which can be explained 
		 taking into account differences between the EuB$_{5.99}$C$_{0.01}$ and the EuB$_6$ \cite{Batkova08}. 
Let us summarize the most essential differences here. 
(i) While the paramagnetic state in EuB$_6$ is homogeneous, the paramagnetic state in EuB$_{5.99}$C$_{0.01}$ 
		is inhomogeneous, containing regions of increased carbon content that are characterized by 
		correspondingly higher electrical conductance in comparison to the remaining matrix \cite{Batkova08}.  
(ii) The magnetic polaron phase in EuB$_6$ can be treated as a two-component system consisting of poorly 
		conductive paramagnetic matrix and highly conductive FM phase represented by MPs \cite{Sullow00a},
		whereas this phase in EuB$_{5.99}$C$_{0.01}$ has three components at least: 
		less conductive regions with lower carbon content, highly conductive FM phase represented by MPs 
		(formed in less conductive regions with lower carbon content), and more conductive carbon-rich domains. 
The latter, incompatible with the existence of MPs due to too high charge carrier concentration,
		 play role of spacers that prevent MPs to link and to form a highly conductive path across the 
		 sample  \cite{Das012,Batkova08}. 
(iii) Finally, magnetically ordered state in EuB$_6$ is a homogeneous ferromagnet \cite{Sullow00a}, 
		 while EuB$_{5.99}$C$_{0.01}$ can be treated as two-component system consisting of FM matrix and HM domains 
		 formed in the carbon-rich regions \cite{Batkova08}.
Here should be mentioned that the mixed magnetic phase 
		 of EuB$_{5.99}$C$_{0.01}$ is expected to be a reason for anomalously high residual resistivity, 
		 which at temperatures as low as 50~mK is even greater than the room temperature resistivity \cite{Batko95}. 

Taking into account the picture of EuB$_{5.99}$C$_{0.01}$ sketched above and the electrical resistivity behaviour, 
		one could expect that fraction of FM phase in this compound near above 7.5~K is still very small, 
		such as volume of this phase is still not sufficient to change semiconducting behaviour of the $\rho(T)$ 
		dependence [d$^2\rho(T)/dT$$<$$0$]. 
(7.5~K is the position of the inflection point in the $\rho(T)$  dependence, as can be seen in Fig.~\ref{G-R0}b). 
However, volume of the FM phase should rapidly increase at lower temperatures approaching $T_M$, 
			and consequently $T_C$. 
Under this assumption,
			the $(I(Q)|_{7~\scriptsize\textnormal{K}}-I(Q)|_{30~\scriptsize\textnormal{K}})/I(Q)|_{30~\scriptsize\textnormal{K}}$
			 dependence shown in Fig.~\ref{SANS} 
			can indicate that scattering at this temperature is a consequence 
(i) of increasing the number of MPs and/or 
(ii) HM domains that are formed in regions of higher carbon content (spacers) already at this temperature.
Because of very similar diagram of scattered intensities at 4~K and 2~K, the latter case is prefered, 
		and only minor amount of small FM-volumes (MPs) is believed to contribute to neutron scattering 
		at this temperature, predominantly at highest $Q$-values. 
As can be seen in the Fig.~\ref{SANS}, 
		maximal relative change in the neutron scattering intensity in the temperature region between 30~K and 7~K 
		is observed for the interval of $Q$ with the centre slightly above 2~nm$^{-1}$. 
This corresponds to neutron scattering on spherical objects with diameter ~3~nm.
Such result is in excellent agreement with our earlier estimation that dimension of the spacers 
			should be around 2.9~nm \cite{Batkova10}. 
Temperature decrease to 6~K is expected to be accompanied by rapidly increasing proportion of the FM phase, 
			because the $\rho(T)$ dependence in the vicinity of 6~K clearly tends to reach maximum, 
			which is believed to be a consequence of formation, grow, and linking of MPs. 
Indeed, such picture adequately explains the rapid increase of scattering intensity in the whole 
			investigated Q-range after cooling the sample from 7~K to 6~K.  
Thus, the relative increase of 
			 neutron   scattering  observed   at   6~K, 
			 $(I(Q)|_{6~\scriptsize\textnormal{K}}-I(Q)|_{30~\scriptsize\textnormal{K}})/I(Q)|_{30~\scriptsize\textnormal{K}}$  
			 dependence, can be associated with volume increase of the FM regions at cooling. 
Such conclusion moreover correlates with the above discussed heat capacity studies indicating high disorder
			 of the system and presence of short range ordering above the temperature of the bulk FM ordering. 
The observed increase of scattering intensities for highest Q (smallest particles) can be associated
			 with the fact that there were formed new MPs, and/or many MPs, which were too small to contribute 
			 to neutron scattering at 7~K have grown in size enough to be detectable at 6~K.
Analogously, according to growing and linking of MPs with decreasing temperature, the MPs detected at 7~K 
			contribute to neutron scattering at highest values of $Q$ than it is at 6~K. 
Further decrease of temperature causes further grow of volumes filled with FM objects, 
			which merge at the temperature of the bulk FM ordering, $T_C = 4.3$~K \cite{Batkova08}. 
It is therefore naturally expected that scattering on individual FM objects should almost vanish at 4~K, 
			and only objects non-compatible with FM ordering (HM domains) should still contribute to scattering. 
Indeed, comparison of $(I(Q)|_{4~\scriptsize\textnormal{K}}-I(Q)|_{30~\scriptsize\textnormal{K}})/I(Q)|_{30~\scriptsize\textnormal{K}}$ 
			dependence with that observed for 6~K and 7~K is in excellent agreement with such a picture. 
Thus the neutron scattering observed at 4~K is believed to be predominantly associated with scattering 
		on HM domains formed in regions with higher carbon content. 
As evident from  Fig.~\ref{SANS}, further cooling down to  2~K does not have any significant impact on
				 $(I(Q)|_T-I(Q)|_{30~\scriptsize\textnormal{K}})/I(Q)|_{30~\scriptsize\textnormal{K}}$ dependence. 
The observed minor changes can be a consequence of minor redistribution of FM and HM phase with decreasing temperature. 

In summary, the obtained Hall resistivity data bring direct proof that strong temperature dependence
		of the electrical resistivity of EuB$_{5.99}$C$_{0.01}$ in the vicinity and close above the temperature 
		of the bulk FM ordering (see Fig.~\ref{G-R0}b) is not a consequence of changes in electron concentration.
As follows from indications for short-range magnetic ordering of performed heat capacity studies 
		and evidence for formation/disapearance of nanometer-sized "objects" at temperatures above and below 
		the  temperature of resistivity maximum as provided by SANS studies, such resistivity behaviour can be 
		adequately explained by disorder of material due to magnetic phase separation. 
In this sense the above presented results bring evidence about mixed magnetic structure in  EuB$_{5.99}$C$_{0.01}$ 
		 and provide another support for our previous proposition that carbon rich regions play role of "spacers", 
		 which are crucial in preventing percolation of MPs and formation of the bulk FM state \cite{Batkova08}. 
Indeed, it seems that just spacers are responsible for high disorder, extremely high residual resistivity, 
			short-range magnetic ordering of the system above the $T_C$, and in general, 
			they are the main reason for huge differences in $\rho(T)$ behaviour between the EuB$_{5.99}$C$_{0.01}$
			 and the stoichiometric EuB$_6$.

\section*{Acknowledgements}
 This research was supported by the Slovak Scientific Agency (Grants No.~VEGA 2-0106-13 and VEGA 2-0184-13), 
 				and by the European Commission under the 7th Framework Programme through the Research Infrastructure 
 				action of the 'Capacities' Programme NMI3-II (Grant number 283883). 
 				M.B and I.B. thank HZB for the allocation of neutron beamtime, and  M.B. acknowledges 
 				also the support of the European Science Foundation (COST-STSM-P16-4334).





\begin{thebibliography}{17}
\expandafter\ifx\csname url\endcsname\relax
  \def\url#1{\texttt{#1}}\fi
\expandafter\ifx\csname urlprefix\endcsname\relax\def\urlprefix{URL }\fi

\bibitem{Degiorgi97}      
L.~Degiorgi, E.~Felder, H.~R. Ott, J.~L. Sarrao, Z.Fisk, Phys. Rev. Lett. 79
  (1997) 5134.

\bibitem{Snow01}      
C.~S. Snow, S.~L. Cooper, D.~P. Young, Z.~Fisk, A. Comment, and J.-P. Ansermet, Phys.
  Rev. B 64 (2001) 174412.


\bibitem{Sullow00a}    
S.~S\"ullow, I.~Prasad, M.~C. Aronson, S.~Bogdanovich, J.~L. Sarrao, Z.~Fisk,
  Phys. Rev. B 62 (2000) 11626.

\bibitem{Calderon04}     
M.~J. Calder\'on, L.~G.~L. Wegener, P.~B. Littlewood, Phys. Rev. B 70 (2004)
  092408.

\bibitem{Yu06}    
U.~Yu, B.~I. Min, Phys. Rev. B 74 (2006) 094413.      

\bibitem{Das012}    
P.~Das, A.~Amyan, J.~Brandenburg, J.~Müller, P.~Xiong, S.~von~Molnár, Z.~Fisk, Phys. Rev. B 86, (2012)  184425. 

\bibitem{Cooley97}     
J.~C. Cooley, M.~C. Aronson, J.~L. Sarrao, Z.~Fisk, Phys. Rev. B 56~(22) (1997)
  14541.

%
%

\bibitem{Aronson99}        
M.~C. Aronson, J.~L. Sarrao, Z.~Fisk, M.~Whitton, B.~L. Brandt, Phys. Rev. B 59
  (1999) 4720.

\bibitem{Kasaya78}     
M.~Kasaya, J.~M. Tarascon, J.~Etourneau, P.~Hagenmuller, Mat. Res. Bull. 13
  (1978) 751.

\bibitem{Molnar81}       
S.~von Molnar, J.~M. Tarascon, J.~Etourneau, J. Appl. Phys. 52 (1981) 2158.

\bibitem{Etourneau85}    
J. Etorneau and P. Hagenmuller, Philos. Mag. B 52 (1985) 589.

\bibitem{Tarascon81}    
J.~M. Tarascon, J.~L. Soubeyroux, J.~Etourneau, R.~Georges, J.~M.~D. Coey,
  O.~Massenet, Solid State Commun. 37 (1981) 133.

\bibitem{Batko95}      
I.~Bat$\!$'ko, M.~Bat$\!$'kov\'a, K.~Flachbart, D.~Macko, Y.~B.~P.
  E.~S.~Konovalova, J. Magnetism and Magn. Mat. 140 (1995) 1177--1178.

\bibitem{Batkova08}     
M.~Batkova, I.~Batko, K.~Flachbart, Z.~Jan\accent23u, K.~Jurek, J.~Kov\'a\v{c},
  M.~Reiffers, V.~Sechovsk\'y, N.~Shitsevalova, E.~\v{S}antav\'a, J.~\v{S}ebek,
  Phys. Rev. B 78 (2008) 224414--1.

\bibitem{Batkova10}   
M.~Batkova, I.~Batko, E.~Bauer, R.T.~Khan, V.B.~Filipov, E.S.~Konovalova, Solid State Commun. 150 (2010) 652.

\bibitem{Hurd72}
C.M.~Hurd, The Hall Effect in Metals and Alloys, Plenum Press, New York, 1972.

\bibitem{Zhang09}   
X. Zhang, L. Yu, S.~von~Moln\'ar, Z.~Fisk, P.~Xiong, Phys. Rev. Lett. 103  (2009) 106602.


\end{thebibliography}
\end{document}